\documentclass[10pt, conference]{IEEEtran}
\IEEEoverridecommandlockouts
\usepackage{cite}
\usepackage{amsmath,amssymb,amsfonts}
\usepackage{algorithmic}
\usepackage{graphicx}
\usepackage{textcomp}
\usepackage{xcolor}
\def\BibTeX{{\rm B\kern-.05em{\sc i\kern-.025em b}\kern-.08em
    T\kern-.1667em\lower.7ex\hbox{E}\kern-.125emX}}
\usepackage{dblfloatfix}
\usepackage{url}
\usepackage{enumitem}
\usepackage{array}
\usepackage{multirow}
\newcommand{\remove}[1]{}    
    
\begin{document}

\title{Collaborative analysis of genomic data: \\vision and challenges
%Genomic data privacy and security: requirements, challenges, existing techniques, and visions
}

\author{\IEEEauthorblockN{Sara Jafarbeiki$^{\ast\dagger}$, Raj Gaire$^{\dagger}$, Amin Sakzad$^{\ast}$, Shabnam Kasra Kermanshahi$^{\ddagger}$, Ron Steinfeld$^{\ast}$}
\IEEEauthorblockA{\textit{$^{\ast}$Faculty of Information Technology, Monash University, Australia} \\
\textit{$^{\dagger}$CSIRO Data61, Australia} \\
\textit{$^{\ddagger}$School of Computing Technologies, RMIT University, Australia} \\
\{sara.jafarbeiki, amin.sakzad, ron.steinfeld\}@monash.edu\\
\{sara.jafarbeiki, raj.gaire\}@data61.csiro.au\\
shabnam.kasra.kermanshahi@rmit.edu.au}
%\and
%\and
}

\maketitle

\begin{abstract}
The cost of DNA sequencing has resulted in a surge of genetic data being utilised to improve scientific research, clinical procedures, and healthcare delivery in recent years. Since the human genome can uniquely identify an individual, this characteristic also raises security and privacy concerns. 
In order to balance the risks and benefits, governance mechanisms including regulatory and ethical controls have been established, which are prone to human errors and create hindrance for collaboration.
Over the past decade, technological methods are also catching up that can support critical discoveries responsibly.
In this paper, we explore regulations and ethical guidelines and propose our visions of secure/private genomic data storage/processing/sharing platforms. Then, we present some available techniques and a conceptual system model that can support our visions. 
Finally, we highlight the open issues that need further investigation. 
% Genome-wide association studies (GWASs), diagnostic testing, and personalised medicine are all benefiting from these improvements. 
% Furthermore, government regulations and ethical bodies demand that healthcare data be kept secure and private.
% To handle these privacy and security requirements and challenges, 
% for genomic data privacy and security 
%After that, this paper shows that showcases our visions. 
%%need to add a system model at the end or erase this one here.
\end{abstract}


\section{Introduction}\label{introduction}
% Point 1 - Genomic data is widespread
Thanks to decreasing cost of sequencing genomes, the acquisition of genomic data are increasing. These data help researchers and clinicians in biomedical research/analysis and clinical procedures. 
This advancement is improving existing healthcare practices such as giving care after an illness and helping to anticipate, prescribe, and prevent illnesses from occurring in the first place.
In other words, these analyses enable the delivery of evidence-based healthcare. 
% -- doesn't fit here - will move
% Therefore, collaboration is needed on genomic dataset to provide benefit to human being.\par
%Computation requirements
%genomic aggregation, GWASs and statistical analysis, sequence comparison and genetic testing.

% Point 2 - Genomic data has specific characteristics
Genomic data have specific characteristics that make them sensitive to concerns of privacy and security. 
% 
% Point 3 - Those characteristics can cause lead to harm
% The availability of genomic data leads to 
% many concerns including its privacy and security. 
This concern is mainly due to special features of genomic data, which include (i) a link between characteristics and specific diseases, (ii) the capability to identify individuals, and (iii) the disclosure of family relationships. 
For example, if a person has mutations in BRCA genes, she is more likely to develop breast cancer \cite{satyananda2021high}. 
Due to the impact of data leakage on personal life \cite{privacy1,privacy2}, including job loss due to medical history, no employment, and excessive insurance premiums, care must be taken at all phases of the genomic data generation, storage, and use. 
Furthermore, the owner's consent for the use and reuse of the data becomes important, and hence the governance of data becomes critical.

% Point 4 - 
Current governance based mechanisms (see Section \ref{requirements}) mostly include policies and legal instruments that rely on human based processes. 
Although these processes work in most situations, they are prone to human errors, 
slowness due to the necessary bureaucracy and 
hindrance for collaboration. 
Technologies can help assure the adequate processes are followed with less errors 
and at the speed that would allow collaborations among researcher to foster towards new discoveries.

There are surveys that cover various techniques to address the privacy and security requirements of genomic data, such as \cite{naveed2015privacy,mohammed2020ensuring,berger2019emerging,aziz2019privacy}. However, there is no work to inclusively, (i) identify the current state of genomic data privacy and security solutions, (ii) see the future visions, and (iii) connect them with the enablers (techniques). Therefore,   
%%%%%%%%%%%%%%%%%%%%%%%%%%%%%%%%%%%%%%%%%%%%%%%%%%In this regard, this paper explores the ethics/regulations requirements. Then, based on the existing barriers, presents the future visions of genomic data privacy and security. Consequently, this paper introduces the potential enablers that can provide us with the visions. It also comes up with the existing gaps/issues that are more challenging and are not entirely solved yet.
%%%%%%%%%%%%%%%%%%%%%%%%%%%%%%%%%%%%%%%%%%%%%%%%%
% Point 5 - We present a vision for future and show how current technologies... and identify challenges.
in this paper, we explore a vision for the future of genome analysis, analyse how technologies can contribute to achieving those visions, and the open issues that still need further research and development. 
While governance has a role in genomic data privacy/security assurance, our vision is not to rely on policies (or non-technical aspects) alone. By putting more effort into techniques, such as cryptography, novel approaches can be developed that can assist humans and guarantee the required security and privacy requirements.
Before getting into those details, let us first look into some background on genomic data to establish a context on the subject matter. 
% DOES THE FIGURE BELOW NEED TO BE AT THE BOTTOM OF THE PAGE?

%In the format, it is written that the fig needs to come after it has been cited. Since I have cited it at the top of the page, the fig needs to come after that.
% GOT IT!
\begin{figure*}[b]
\centering
%\vspace*{-1.6cm}
\includegraphics[scale=0.9]{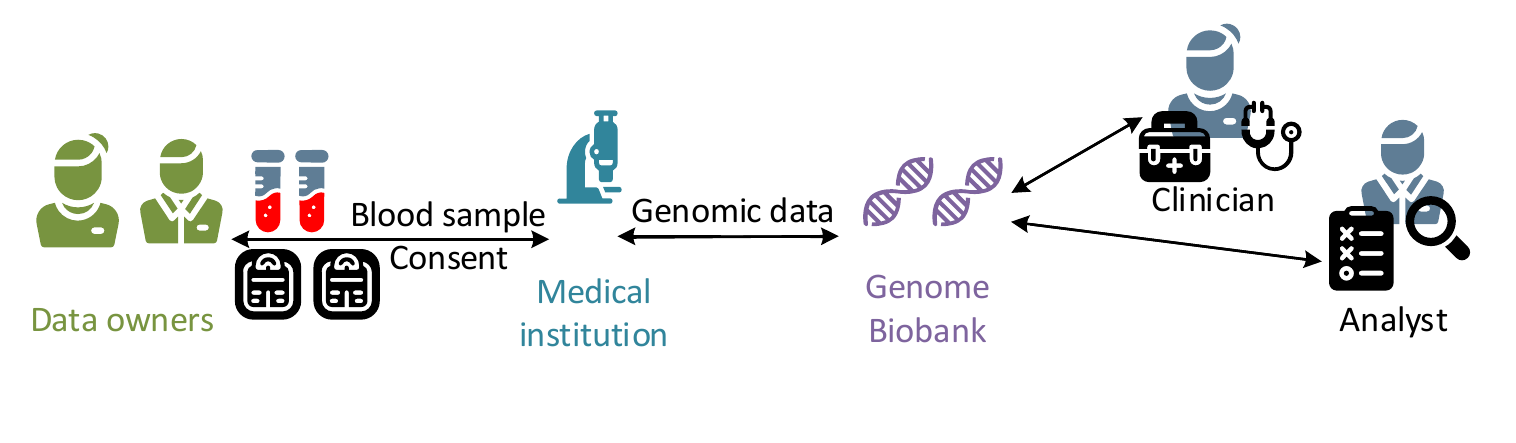}
\caption{Genomic data from generation to use in medical analysis/care}
\label{fig:system design overview}
%\vspace{0.2cm}
\end{figure*}
% how is it collected

% Organisation of this paper...

% BACKGROUND
% \section{Background}

% Genomic data
\subsection{Genomic data and its characteristics}
The genome shows an organism's overall genetic information, encoded in deoxyribonucleic acid molecules know as DNA. DNA consists of two long complementary polymer chains of four units called nucleotides. These nucleotides are represented by the acronyms A, C, G, T (Adenine, Cytosine, Guanine and Thymine). It is a chain-like molecule of variable length made of these letters.
It includes details on a person's health and behavior. It does not change much over time. 
Although most of the human genome look alike, small changes across the large genome make it unique for each individual. 
That means any two people's DNA can be easily distinguished from one another. 
Moreover, an individual's DNA carries information about her biological relatives and leaks information beyond the individual from whom it was acquired \cite{hubaux2017genomic}. 
It can be used to recognise disease predisposition and physical traits that make DNA a large collection of sensitive information. Such knowledge about a person can have significant implications \cite{lin2004genomic, li2019robust} and 
is vulnerable to privacy risks and attacks \cite{naveed2015privacy}. %Because DNA involves kinship, it leaks information beyond the individual from whom it was acquired.
%Several possible correlation attacks within genetic data and with other data (phenotypic, or even data obtained from online social networks) have already been demonstrated, with many more to come.

\subsection{Genomic data collection process}
The process of genomic data collection is shown in Fig. \ref{fig:system design overview}. %\footnote{AS: Fig.~1 needs to be updated, which one is clinician? Which one is analyst?} 
As illustrated, patients and also healthy people (as data owners) provide their samples (usually blood samples) to a medical institution (or hospital or a gene trustee). 
They also provide consent for different uses of their samples, including the usage of information available in their samples. 
The medical institutions collect, sequence (i.e., determine characters of DNA) and process (i.e., clean up and prepare for further analysis) the genomic data. 
Then, they manage the information (genomic data, diseases, metadata such as gender, ethnicity, age and consent) and store the data in a biobank. 
For meaningful analysis, researchers need to test their hypotheses on a large number of samples. Therefore, these data are shared in public databases, biobanks and repositories (e.g., UK biobank \cite{UKbiobank} and the 1000 Genomes Project \cite{1000genomes}).

% how is it used
\subsection{Uses of genomic data}
There are several usage of genomic data \cite{naveed2015privacy}. Here, we focus on two main usages: personalised health and disease research.
% This section discusses the major applications of genomic data that are related to our visions.\\%\footnote{AS: Are these the only important applications?}\\

\textbf{Personalised Health}: It has long been known that changes in a person's genetic sequence can have an impact on their health. Changes in a particular gene can have a negative impact on a person's health instantly or in the future \cite{botstein2003discovering}. While some of these diseases are treatable with dietary modifications or pharmaceutical therapies, others are not, and there is no proven intervention to help improve a person's health. Nonetheless, some people want to know their genetic status to prepare for the expected future and contribute to medical research.
Genetic testing has applications in areas such as (i) paternity testing, to decide if two individuals are parents and offspring, (ii) diagnostic testing, to determine if an individual is impaired by a specific disease and (iii) pharmacogenetic testing, in personalised medicine to customise patient care and to develop successful drugs and therapies \cite{yakubu2020ensuring}.

\textbf{Disease Research}: Despite the fact that the genome has been connected to a large number of diseases and treatment responses, new connections are being identified regularly.
The technology for doing basic research is rapidly evolving \cite{logsdon2020long}. 
Today's amount of genetic data being gathered, stored, and processed is extraordinary, hastening the identification of new mutations.
Meaningful discoveries require a large number of relevant samples to ensure the statistical significance of the results.
Since gathering such a large number of samples in a single institution may not be possible, the data need to be shared among researchers in different institutes.
As more collaboration is needed on genomic data for research, so too will the privacy and security assurance of genomic data \cite{bonomi2020privacy}. 
% Therefore, people will engage and provide their samples for more research.

% \remove{\textbf{Legal and forensics}: A suspicious record in a forensic database has to be checked against the whole database.
% DNA collected at the crime site is often examined in order to hunt down the perpetrator. In certain nations, it is permissible to collect and preserve a suspect's DNA for this purpose.\par}
Due to the sensitivities of these data, several governance mechanisms, including legal and ethical instruments, have been established to ensure the data is shared and used in a responsible way.

\section{Governance of genomic data}\label{requirements}
The genome data has special features and cannot be treated like other data. 
Hence, it is important to consider protective measures before sharing the genomic data, to provide privacy of the individuals who contribute the data and reduce harm from the data.
Adhering to the law and ethics in one of such important considerations.

% DOES THE TABLE BELOW NEED TO BE AT THE BOTTOM OF THE PAGE?
\begin{table*}[b]
\caption{Overall requirements due to law/regulations}
\begin{center}
\begin{tabular}{|>{\arraybackslash} m{4cm}|>{\arraybackslash} m{4cm}|>{\arraybackslash} m{4cm}|>{\arraybackslash} m{4cm}|}
\hline
\centering\textbf{Overall}&\multicolumn{3}{|c|}{\textbf{Regulations}} \\
\cline{2-4} 
\centering\textbf{requirements} & \centering\textbf{\textit{Privacy Act}} \cite{privacyact}& \centering\textbf{\textit{GDPR}} \cite{GDPR}& \multicolumn{1}{|c|}{\textbf{\textit{HIPAA}} \cite{HIPAA}}  \\\hline
Confidentiality, integrity and availability of health data&Security and integrity are discussed. &Data
security and end-to-end
encryption are considered. %data protection bydesign and by default. 
& More Health data must be kept
confidential, integral, and
available. \\\hline
Transparency and fairness&Open and transparent handling of personal information are required, as well as anonymity and pseudonymity.  &Lawfulness, transparency and fairness are considered when handling personal health data. &Policies, methods, and technology that directly influence individuals and their health information should be open and transparent.$^{\mathrm{a}}$ \\\hline
Consent&Consent is required for disclosing personal health data and it can be only used for that purpose consent has been given for. Consent can be withdrawn by individuals. &The aim of data use must be clearly stated, unambiguous, and legal. The health data has to be used for reasons for which consent (can be withdrawn) was collected. &Consent is required for secondary use of health data, and in a written form. Consent can be revoked by individuals in writing.\\\hline
Access control&Individual access can be provided on request. &Protect health data against unauthorised or unlawful processing and accidental loss or damage. &Access has to be authorised, e.g., considering access control with unique user identification. \\\hline
Breach notification&Affected individuals and the Australian Information Commissioner have to be notified about a breach. &Individuals have right to be notified if their data is breached within 72
hours. &Provide notification following a breach of an individual's personal health data.
 \\
\hline

\multicolumn{4}{l}{$^{\mathrm{a}}$The U.S. Department of Health and Human Services (HHS) reformed the requirements \cite{baker2016governance}.}
\end{tabular}
\label{table:law}
\end{center}
\end{table*}

\subsection{Law and Regulations}
The law is a set of rules that a country or society acknowledges as governing its people's activities and that it may enforce through the application of penalties \cite{law}. It is governed by a government. Ethics, on the other hand, are moral rules that regulate a person's behavior or active conduct \cite{ethics}.
%Some of the selected requirements that have been documented from governments and institutions based on legal and ethical concerns are discussed in this section. The sources that discuss the major requirements due to law and ethics for the related field, health data specifically genomic data, are considered here. To more clarify/explore the consent, it has been separately discussed at the end of this section.
% \subsection{Governance/law}%legal concerns, Regulations,Governance/law,Legal requirements
Several nations and organizations have taken major steps in enacting laws and regulations governing health data privacy, security, and trust.

The Privacy Act 1988 \cite{privacyact} governs the privacy and security framework for personal information in Australia. Most Commonwealth government agencies (including the tax office and the Department of Human Services), all private sector organizations with an annual turnover of more than three million dollars, and a few other organizations that meet specific criteria, such as health service providers, are covered by the Privacy Act. The Privacy Act 1988 (‘Privacy Act’) defines ‘health information’ as including ‘genetic information about an individual in a form that is, or could be, predictive of the health of the individual or a genetic relative of the individual’.

GINA (Genetic Information Nondiscrimination Act of 2008) \cite{GINA} and, more subsequently, HIPAA (Health Insurance Portability and Accountability Act of 1996) \cite{HIPAA} are laws that deal with genetic information. GINA is an anti-discrimination law. It prohibits group health and Medicare supplementary plans from using genetic information to discriminate against individuals when it comes to insurance, but not life, disability, or long-term care policies. It also prohibits businesses from utilising genetic data in hiring, firing, promotions, or work assignments.
Similarly, HIPAA provides guidelines 
about how to handle sensitive information through de-identification. HIPAA laws were updated in 2013 by the HIPAA Omnibus Rule, which included genetic information in the scope of Protected Health Information (PHI).

The General Data Protection Regulation (GDPR) \cite{GDPR}, a new framework for European data privacy legislation, was authorised and enacted by the European Commission in 2016. The GDPR protects individuals' privacy rights and aims to harmonize data protection rules across Europe, regardless of where data is handled. The GDPR and genomic data report \cite{GDPRreport} examines how the GDPR affects genomic healthcare and research, highlights areas that require attention, and gives recommendations to the genomics community.

% how is GDPR related to genomic data? You might want to add a couple of sentences about its relevance.
%The global alliance for genomics and health (GA4GH), that creates policies and standards for the secure sharing of genomic data.
Some of the overall requirements due to these regulations are discussed in Table \ref{table:law}.

\subsection{Ethics}%ethical concerns, Ethics,Ethical requirements
%\footnote{Why only India and Australia? Other countries does not have it or what? Can you please elaborate?}
In Australia, the laws governing genetic data exchange are a mix of common law, legislation, ethical principles, and codes of practice. The primary regulatory lever is the National Health and Medical Research Council's (NHMRC) National Statement for Ethical Conduct in Human Research (National Statement) \cite{nhmrc}, which applies to genetic data that satisfies the criteria of personal information.

The Indian Council of Medical Research (ICMR) \cite{icmr} has provided comprehensive guidelines for Indian researchers working in the genomic area, and a separate chapter has been included in the 2006 issue of “Ethical guidelines for biomedical research on human subjects.”
The prospect of re-identification of users through connecting, combining, data-mining, and reusing large databases raises ethical concerns. 

There are other ethical studies such as \cite{italyethic} which is performed by an Italian company named Evodevo srl, and is supported by the European Economic and Social Committee. It explores the ethical requirements of big data, which also includes health and genomic data. Another ethics guideline can be found in \cite{USAethic}, which is based on the American Medical Association code of medical ethics.

In various ethical guidelines such as the above-mentioned ones, the main points are (i) privacy and confidentiality, and (ii) consent. Apart from these two, ethics approval in research projects, data sharing management, and risk analysis and management are also discussed.

Legal and ethical aspects in genomic data security and privacy provide some ways of data protection.
%The Health Insurance Portability and Accountability Act of 1996 (HIPAA) establishes rules for the de-identification of sensitive information.\par
Different health data security/privacy issues have been solved using policies/regulations.
For example, \emph{consent} is taken from data owners for different uses of their genomic data using papers with all the information included to be signed. This does not resolve different requirements of consent revocation/update thoroughly. Moreover, when using genomic data for analysis to \emph{generate knowledge}, the dataset is being downloaded by authorised researchers/analysts. The problem is that the consent revocation is hard in this case and after downloading data, it would be nontransparent. Furthermore, for more accurate analysis, when data is stored in different organisations, \emph{interconnection and integration} becomes a key. It is done in an institutional way, where a consortium is created and people signed up to that consortium are able to share data and access it. In addition to all the above, \emph{transparency and accountability} are essential when working with sensitive genomic data. The data is distributed to analysts based on policies/regulations and penalties are enforced to someone who does not obey them.

% PROBLEM
\section{Problems}
% --- how big the problem is, how important the problem is

By relying only on regulations and ethics to provide guidelines, some problems arise that are explained in this section. 
% Human Error
\subsection{Human error}
When human is involved in the genomic data handling and process, human error can cause problems. 
Example scenarios of human errors may include (i) Sending sensitive information to the wrong recipients; (ii) Sharing password/account information; and (iii) End user and IT staff failure to follow the rules and procedures. Therefore, sensitive data is exposed as a result of data breaches \cite{DataBreach101, DataBreachcomptia}.
The healthcare business is one of the worst affected in comparison to other data industries \cite{liu2015data}.
According to several practitioners, the overall number of people affected by healthcare data breaches was $249.09$ million from 2005 to 2019 \cite{DataBreaches}. Human error was also identified as a top source of Australian data breaches in 2020, where it accounted for $38\%$ of the breaches \cite{AUDataBreaches} and health industries was one of the most affected industries.
If a breach occurs, genetic data may be used to identify an individual and forecast their physical traits \cite{lin2004genomic, li2019robust}, and familial matching can be performed by matching individuals to distant relatives \cite{erlich2018identity}. Furthermore, individual and family health information can be revealed using genetic data \cite{sawaya2020artificial}.

\subsection{Trust-based mechanisms}
%When data is shared with the users, not distributing the data would be based on trust and penalties.
Upon approval, an analyst can 
% Based on a guideline, whenever a request form is completed through an analyst to get access to the data, and she is able to 
download the requested data, and it will be her responsibility if misused. This process is being handled based on trusting the authorised users not to distribute the data or misuse it. Otherwise, penalties might be applied. However, apart from the trust requirement that can be an issue, human error may be added when data is downloaded and used. This situation leads to the data breaches mentioned in the previous subsection.

% Hindrance to Collaboration
\subsection{Hindrance to collaboration}
%availability of data
%paperworks, governance processes
Human involvement can make a process slow. Relying on the human to assess the requests and authorise access needs governance processes and paperwork that needs time to complete. As such, data requests may take a long time to process. This bureaucracy lowers the speed of analysis.
% , which is related to the availability of the data. 
If the data analysis process can be automated and all the checks can be done using computational techniques, not only will the process be quicker, but also the human error in authorising and assessing will be eliminated.

% Consent - 
\subsection{Consent management}
%For example, in taking consent and revoking it, some errors may lead to not acting accordingly and missing some revocation requests.
%one time consent, written requests
% --
An informed consent is required in most conditions when a medical procedure is performed. %E-Consent is a model solution that enables patients to have greater control over who can access their information. Although individuals usually provide unconditional consent to access their details, %This allows organizations to request information in the future when such requests are unwarranted. %E-Consent is a broad term, specific forms can be categorized as follows [15]:
%General consent grants blanket permission to access information. This may become problematic if patients move, change physicians, or only perform a singular procedure at a site. Furthermore, there is no limit to what data can be requested.
%%%%%consent is taken as a matter of legal record. %E-Consent systems may be more involved by encouraging clinicians to confirm that they have acknowledged conditions of consent prior to accessing a record. Finally, the patient consent record could be completely active and used as a gatekeeper \cite{coiera2004consent}.
%In other words, it is typically the duty of people working in the health care system to obtain consent from a patient or to assess if consent exists, before obtaining, using, or forwarding the information of a patient.
For using the data in analysis and study, to obtain consent, the subject must be informed of his or her rights, the intent of the study, the procedures to be performed, the possible risks and benefits of participation, the anticipated length of the research, the degree to which personal identity and demographic data are confidential, so that subjects' participation in the study is entirely informed and voluntary \cite{nijhawan2013informed}. However, one-time consent is usually not sufficient and individuals should be able to revoke their consent regarding the access to their data.

In general, consent may be approached in two ways: static consent and dynamic consent. Static consent requires that an agreement, generally a paper document, is signed by the participants at the moment of data collection for all future data uses. This approach does not consider the difficulties that arise as the environment and requirements change over time, such as utilising the data for a different health initiative than the one for which the consent was given.
Furthermore, written consent forms are hard to manage, especially when it comes to their revocation. In addition, when the data is shared with analysts and a revocation request comes in, managing it would be a problem. Automating the consent (that goes with the data) and managing it efficiently is still a problem.

% --- how big the problem is, how important the problem is
% Human Error

% Hindrance to Collaboration

% Consent - 

% \remove{
% Some of the overall requirements due to these regulations are as follows:
% \begin{enumerate}
%     \item Confidentiality, integrity and availability of health data
%     \item Secure and privacy-preserving health data processing
%     \item Transparency and fairness
%     \item Consent
%     \item Access control
%     \item Breach notification
% \end{enumerate}\footnote{AS: can you put all these in a table vs all the above mentioned acts/regulations and say which one addresses which?}}

% --

% TECHNOLOGIES TO SUPPORT THE VISION

% SYSTEM MODEL THAT INCORPORATES THE VISION

% OPEN ISSUES AND CHALLENGES
% --- Trade-off of security/privacy vs usability

% CONCLUSION

% REFERENCES

% -- Explain some of the concerns here

% \section{Paragraphs to move elsewhere}
%However, traditional methods of genomic data privacy protection, such as de-identification, are insufficient.

In light of these existing issues, it is critical to investigate the requirements/techniques for genomic data privacy and security. Considering the existing problems, we discuss our future visions in the next section. Some of our visions are probably near future visions, and some need more investigations and might take time to be established.

%Collaboration requirements-Cancer example

%Access to data
%rules, you would spread it

%Consent go with the data

%In the healthcare domain, the human genome is special and unique, it does not alter much during an individual's life, it represents ethnic heritage, it correlates with relatives and can recognise disease predisposition that makes it a large collection of sensitive information and is vulnerable to privacy risks and attacks \cite{naveed2015privacy}.

% \section{Genomic data privacy and security requirements}\label{requirements}%concerns
%\subsection{Data management}%privacy risks/attacks

% \section{Vision for Future}
% VISION FOR FUTURE

\section{Future visions}%vision for future
\begin{figure}[htp]
\centering
%\vspace*{-1.6cm}
\includegraphics[scale=0.68]{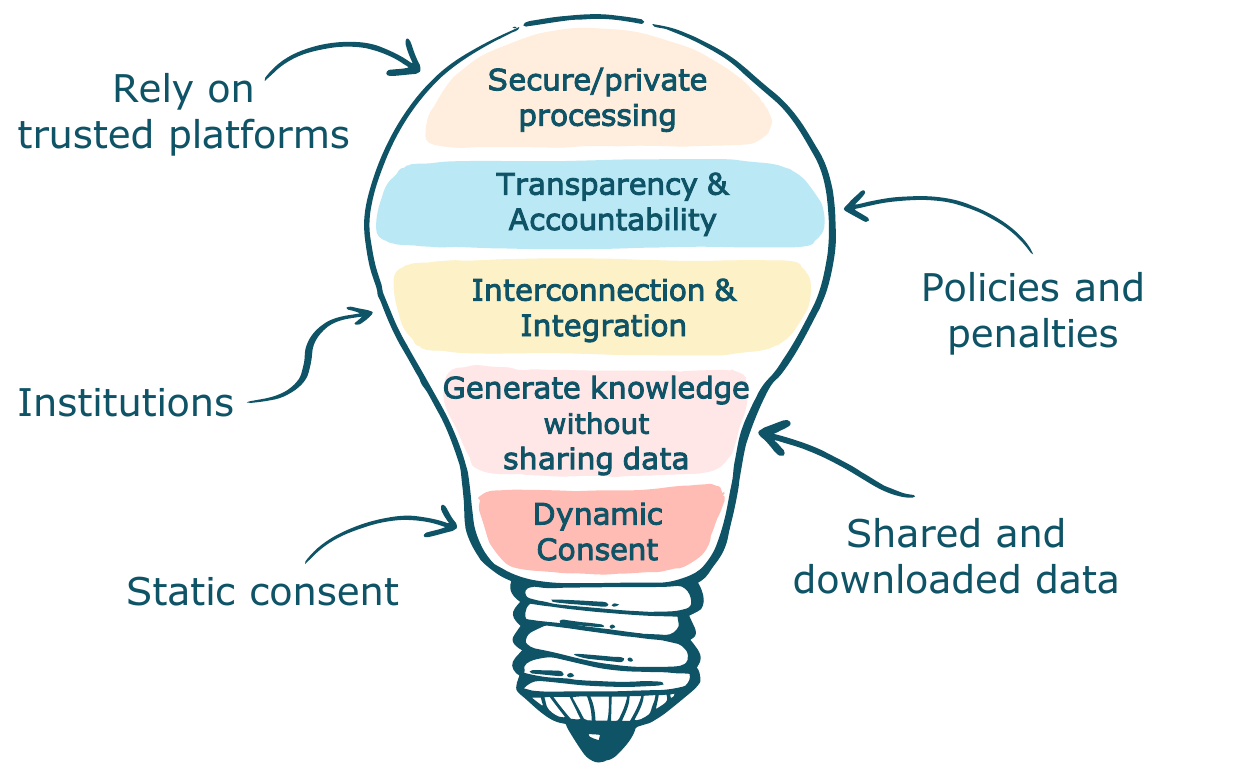}
\caption{Future visions}
\label{fig:vision}
%\vspace{0.2cm}
\end{figure}

This section presents our visions with respect to genomic data privacy and security. The future visions and their current states that have some issues (discussed at the end of section \ref{requirements}) are represented in Fig. \ref{fig:vision}.%\footnote{AS: inside Figure, Dynamin should change to Dynamic.}

\remove{\definecolor{azure(colorwheel)}{rgb}{0.0, 0.5, 1.0}
{\scriptsize\fcolorbox{azure(colorwheel)}{azure(colorwheel)}{\rule{0pt}{2pt}\normalfont{\textbf{\textcolor{white}{1}}}\rule{1pt}{0pt}}} 
}

\subsection{Secure and private processing/computing of genomic data}
Computation on genomic data is an important factor in the data analysis to help researchers with statistical analysis, genome-wide association studies (GWAS), etc. %\footnote{AS: do what?}. 
There are platforms such as Google cloud confidential computing and Azure confidential computing %\footnote{AS: such as? with reference please.} 
that make it convenient, but this way, there is a need for a trusted platform.
This vision can also be combined with the fourth vision, which is \emph{generate knowledge without sharing data}. Therefore, our vision is that while the data is not shared with requesters, the computations are done in a secure/private way on the data. Different techniques have been used to provide this in the literature but they either rely on a trusted platform (e.g., Google cloud) or their protocols (e.g., Homomorphic encryption-discussed in section \ref{techniques}) are not ready as a product to be used.
%We considered our future vision as the knowledge being generated without the data being shared. 
In this regard, we present our vision as there has to be computations/analysis done on the stored data that is not shared with anyone and in a secure/private way.%\footnote{AS: I suggest you move the above two sentences to the end of this subsection.}

\subsection{Transparency and accountability}
Accountability requires transparency. %\remove{It is difficult to assess whether data is equitably and effectively provided if data stewards are not forthcoming about data availability and access methods.\footnote{AS: did not understand this sentence, what do you mean stewards are not forthcoming?}}
To build trust, governance of genomic data must be based on transparency and accountability.
It is required to maintain an accounting of all disclosures, even including those related to treatment and healthcare operations.
It is difficult to determine if data users are accountable for the data entrusted to them if they do not make efforts to verify that usage limits are observed. Clear data exchange is required to guarantee stakeholder responsibility. Our vision is to provide technical ways that can support transparency and accountability without just relying on the regulations/guidelines and penalties for those who do not obey the regulations/guidelines and, e.g., share the data with others or misuse it.

\subsection{Interconnection and integration}
There are different institutions/medical centers that might have an individual's medical records/genomic information. If all the data is not in one place there would be some challenges in both analysing the data, and accessing it for medical purposes.
Regarding genome analysis, if an analyst is analysing the data, they should be able to connect them and use them as big data for better output (in other words, integrate data). When it comes to the interconnection of the medical records and genomic data, the challenge goes to the interconnection of data (genomic, medical and health) that are dispersed.
There are relations between genomic data and medical records and health conditions, that need to be taken into consideration while analysing or caring.
Therefore, integration is much needed when the genomic data is stored in different places/hospitals/institutions. More importantly, interconnection is needed when the genomic data is stored in one institution and medical data is stored in another institution or hospital.
Our vision is to integrate genomic data in a dispersed scenario, where genomic data is stored in different institutions and interconnect the genomic data and medical records/health conditions stored in different places but all in a secure/private way.  
%\footnote{AS: what is our vision here?}\par
%The data has to be clean with a standard format.

%\footnote{AS: I suggest all vision statements in all subsections be either in passive or active verbs/voice. I like the one in this subsection.}
\subsection{Generate knowledge without sharing data}
\remove{Based on the need for generating knowledge, either conducting a query on data or performing computations over genomic data becomes challenging when it comes to not sharing the data and performing all the tasks on the data and just revealing the requested information. It is worth mentioning that even the knowledge/information that is generated needs to be checked based on the access the particular requester has or is allowed to get.}

Knowledge can be a specific information about a patient, the result of statistical analysis on genomic data, trained model on genomic dataset, or result of other analysis/queries on genomic data. Genomic data needs to be analysed with other information such as diseases, specifications, e.g., age. Therefore genomic data alongside other information about individuals are gathered and stored.
There are two main ways of storing genomic data, centralised and decentralised storage. In centralised storage, all the data from different hospitals or institutions are collected and stored in one data server. In decentralised storage, the data are stored in different data servers.
It means that there is distributed storage of genomic data, e.g., hospitals. The centralised storage has a risk of failure and attack on all the data we have. However, in decentralised storage, these risks are localised, but computation and processing of data become more complicated.
With accepting the risks in both cases and providing the privacy and security requirements of both types of storage, our vision is to generate all the knowledge needed without sharing the data with analysts. It means that the request comes to the data, the result/knowledge will be generated and sent back to the requester without letting them downloading all the data.
%\footnote{AS: In this subsection you are talking as if a general reader knows about query and the model you have worked with in your research. But you have not talked about this model prior to this subsection. So this has to be a bit more generic.}

\subsection{Dynamic consent}
Owing to the extremely diverse nature of current healthcare settings, one-time consent management approaches are insufficient to address the specific privacy issues and needs of the individual patient \cite{o2013non}.
Dynamic consent is an informed and individualised consent in which the data subject and data custodian have two-way communication and the subject may update and grant various types of consent. In addition, the subject may use the interface to oversee their health data usage over time and revoke consent. Furthermore, the consent is transmitted with the related data when it is shared with third parties, and the participant has access to the research findings. Dynamic consent, on the other hand, has challenges, such as higher implementation costs, consent revocation, and data deletion assurance, as well as the requirement that patients have adequate digital expertise and time.
Some works have already been done on consent to model the consent and manage it such as \cite{IBM} that models the consent in a tool, have a repository for storing it, and a data access management feature to enforce consent. Dynamic consent management has not been implemented and used widely in hospitals/institutions/gene trustees. It is our vision that dynamic consent will be designed and used widely (while providing data security and privacy) to ensure data owners have full control over their data.

\remove{\subsection{Decentralisation}
This can be categorised in the second vision, generate knowledge without sharing data. However, we discuss it here as it can be a further vision.
In practice, the health-related data is decentralised. Different groups' data reside in different hospitals/institutions' storage, or as a vision, each individual's data can be stored on their device.}

\section{Techniques to support visions}\label{techniques}
\begin{figure*}[btp]
\centering
%\vspace*{-1.6cm}
\includegraphics[scale=0.35]{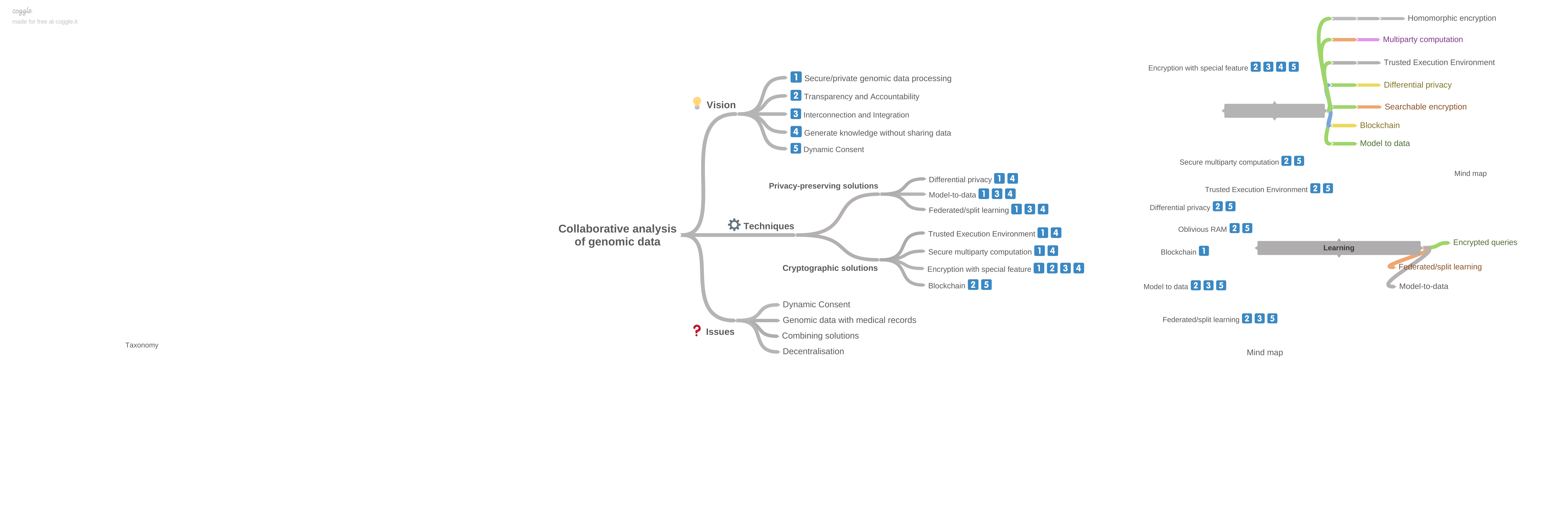}
\caption{A taxonomy of future visions, existing techniques to support visions, and open issues for genomic data privacy and security}
\label{fig:mindmap}
%\vspace{0.2cm}
\end{figure*}

In this section, we introduce some important data security
and privacy-preserving techniques that are able to support our visions. %\footnote{AS: can you please prepare a table showing which techniques can support which vision?} 
%\footnote{AS: Maybe divide this section into to subsections. One on cryptographic solutions and the other one on privay-preserving ML-based solutions.}
A taxonomy of future visions, existing techniques to support visions, and open issues for genomic data privacy and security has been shown in Fig. \ref{fig:mindmap}. The relation between techniques and visions has been demonstrated by using numbers in the taxonomy.
It is worth mentioning that there may be other techniques that can assist with reaching future visions, and also these techniques can be combined and used together in some cases to provide the requirements.

\subsection{Privacy Preserving solutions}

\subsubsection{Differential privacy} 
Differential Privacy (DP) protects the privacy of output data or results from computations or processes by limiting the amount of leakage of individual input data to the authorized (usually minimal) amount. The Laplace mechanism \cite{dwork2006calibrating} and the exponential mechanism \cite{mcsherry2007mechanism} are the most common methods for realising differentially private algorithms, in which a random noise generated from a Laplace distribution and a scaled symmetric exponential distribution are added to the output data to achieve DP, respectively.
Local and global differential privacy are the two forms of DP.
Each dispersed participant adds noise to their input data before collection or computation in local DP, whereas noise is added to the final output after computation in global DP.
DP is used in conjunction with an encryption method. In a dispersed clinical setting, a combination of encryption and DP is utilised to ensure the privacy of genomics data \cite{raisaro2018m}.

% --- [TODO] add other privacy perserving techniques? 
\subsubsection{Model-to-data approach}
Machine learning (ML) approaches are the important part of health data computation. Based on data accessibility, there are two sorts of computing techniques for the stored health data, data-to-modeler (DTM) and model-to-data (MTD) \cite{guinney2018alternative}.
A data modeler has direct access to the data for model building (training and validation) and hypothesis testing in the Data-to-Modeler (DTM) method. Because the data modeler must be trusted, and data, which is sensitive in the healthcare sector, is directly accessible by the modeler, this technique does not follow the privacy rules since the first phases of relevant design and creation (privacy-by-design approach).
The data is not directly accessible in Model-to-Data (MTD); thus a modeler must submit their codes and models (derived from their initial data) to the data provider. To put it another way, the model moves to the data. The models are then trained and verified on the unseen actual data at the data contributors' server. The modeler is notified of the revised model or outcome.
MTD takes a privacy-by-design approach. It allows for computation without access to the personal information \cite{thapa2020precision}.\par %\cite{thapa2020precision}.
%\footnote{AS: are these used to support any of our visions?}

\subsubsection{Federated and Split learning}
Federated learning (FL) and split learning (SL) are distributed deep learning algorithms that do not necessitate the exchange of raw data from multiple locations.
FL shifts computation to edge devices. It protects the privacy of data at all participating edge devices. The training data are normally provided in one central server (or datacenter) in the classic ML technique, and the training takes place on that server, which is generally a trusted platform that can see data while computing. The FL method, on the other hand, protects the privacy of data stored at edge devices by undertaking collaborative learning that never needs data from dispersed edge devices to be acquired outside of those devices \cite{bonawitz2019towards}.
SL is particularly beneficial when data sharing across data holders/sources is impossible owing to resource constraints, as well as privacy and legal concerns. To begin, each client trains a neural network up to a certain layer, called the cut layer, and then sends the activation data from that layer to the server. The server then trains the remainder of the network layers and completes the forward propagation after receiving the activation from the client. Second, the server computes the loss using the true and predicted labels, then backpropagates the loss to calculate the correction in the form of gradients until the cut layer is reached. After that, the cut layer's gradient is sent back to the client. The client continues the network's backpropagation on its side after receiving the gradients. The forward and backward propagations will continue in this manner until the network has been trained \cite{vepakomma2018split}.\par
Federated learning has been used in healthcare applications such as \cite{nvidia}, that employs FL on brain tumor segmentation data. It has also been used to predict hospitalisation of patients who have heart diseases \cite{brisimi2018federated}. It has also been discussed in \cite{makarious2021genoml} as an expansion of their project, which is automating machine learning workflows for genomics, to enable learning across different genomic data storages.

\subsection{Cryptographic solutions}

\subsubsection{Trusted execution environment} The Trusted Execution Environment (TEE) protects important computations by isolating them from other processes like as operating systems, BIOS, and hypervisor. Furthermore, by using isolation and cryptography to minimize the attack surface, it improves the security of the processes occurring in TEE. For the secrecy and integrity of data and application code, TEE employs either a hardware or software module, or both.
Intel Software Guard Extensions (SGX) \cite{costan2016intel} is one of the notable TEEs, that are instruction-set architectural extensions that use trusted hardware to offer a secure computing environment. A privacy-preserving international cooperation platform for analysing rare disease genomic data has been proposed in \cite{chen2017princess}. Intel SGX is used in this study to do secure calculations on dispersed and encrypted genomics data.
%\subsection{Searchable encryption}

\subsubsection{Secure multiparty computation} Multiparty Computation (MPC) allows distributed computations on encrypted data without the need for decryption. Each data input is split into two or more shares, which are then distributed among several (untrustworthy) parties. Without exposing their inputs to the other parties, all parties follow a procedure and jointly calculate a function on their inputs. They are all given a copy of the final result. It is not necessary in MPC to keep all data from various parties centrally, which would need the use of a trustworthy third party.\par
MPC has a lower computational cost in comparison to Homomorphic encryption (discussed in the next subsection), but it has a high communication cost. It has to be considered that MPC alone does not provide the output privacy, as the final result may leak some information about the data. Therefore, it can be used in combination with differential privacy. To construct secure MPC protocols, homomorphic encryption \cite{damgaard2012multiparty}, garbled circuits \cite{yao1982protocols}, linear secret sharing \cite{shamir1979share}, and Oblivious Random Access Machine techniques \cite{gentry2013optimizing} have been utilised. MPC has been used in conducting queries on genomic data to generate knowledge in \cite{hasan2018secure}. It has also been used in \cite{cho2018secure} to provide secure genome-wide association analysis.

%\subsection{Genomic data protection}
\subsubsection{Encryption with special feature} 
There are several encryption mechanisms with special features such as homomorphic encryption \cite{dowlin2017manual}, searchable encryption \cite{zhang2017searchable}, attribute based encryption \cite{li2012scalable}, identity-based encryption (IBE) \cite{chatterjee2011identity}, proxy re-encryption \cite{bhatia2020towards}, order-preserving encryption (OPE) \cite{ayday2013privacy}, etc. that can also help the visions. 
We limit our discussion to the first three methods to provide the readers an overall sense of the potential utility of these techniques.

\paragraph{Homomorphic Encryption (HE)} 
It permits calculations (arbitrary functions) to be performed on encrypted data without the need to decrypt it. The data and results, which are both encrypted, would be unknown to the computing environment. As a result, HE makes it possible to do safe computation on an untrustworthy computing platform. HE requires high computations and overhead, which makes it challenging to implement in real-world applications. Some optimisations have been done, but in the general case, it is still insufficient. HE has been used in literature for private computation on genomic data \cite{lauter2014private}, secure testing for genetic diseases on encrypted genomes \cite{zhou2018secure}, and achieve genome-wide association study in \cite{gwaspaper}, which securely and privately examines genetic variations and single-nucleotide polymorphisms of genetic data.

\paragraph{Searchable Encryption (SE)} 
Classical encryption of data may resolve growing concerns about the protection of outsourced data. It is not simple in reality, though, because the cloud server cannot search the encrypted data directly. Therefore, the user must download all the information and conduct the search after decrypting the data. Searchable Encryption (SE) can solve this problem; it allows the data owner to store data in encrypted form on the cloud while maintaining the server's ability to search through encrypted data.
    SE has been used to let analysts/clinicians execute queries on genomic data without decrypting it \cite{PrivGenDB}. It has also been studied to protect genomic data when enabling substring search \cite{substringsearch}, or range query \cite{rangequery} over encrypted genomic data sequences.

\paragraph{Attribute-based encryption (ABE)} It is a special type of encryption in which the user's secret key and the encrypted data are dependent on the attributes (such as location, job, specialty, etc.). The encrypted data can only be decrypted if the user key's set of attributes matches the ciphertext's set of attributes. It has two types: key-policy attribute-based encryption (KP-ABE) \cite{KP-ABE} and ciphertext-policy attribute-based encryption (CP-ABE) \cite{CP-ABE}.
    CP-ABE has been used along with HE in \cite{namazi2019dynamic} to provide access control based on the attributes of the users of the system while they request for tests/computation on the genomic data.
% \end{itemize}

% However, we do not discuss all the existing mechanisms in this paper.

\subsubsection{Blockchain}
A distributed public ledger, or blockchain, is a chain of blocks that keeps track of transactions or data in a sequentially growing list. The data included in the blocks are both immutable (no one party has the ability to remove it) and time-stamped. Blockchain facilitates data exchange without relying on a network's compute nodes, and it is not controlled by a central node. The data integrity of blockchain allows the security of networks and systems.
Blockchain is utilised in conjunction with homomorphic encryption and secure multi-party computing to provide end-to-end secrecy of genomic data queries \cite{grishin2021citizen}. 
Blockchain is also studied to provide dynamic consent framework \cite{mamo2020dwarna}.
Despite the possibility to increase data security by blockchain data encryption, there is the risk of data leaking from a public blockchain as a result of attacks, such as linkage attacks \cite{agbo2019blockchain}.

% \subsection{Privacy-preserving machine learning (ML)-based solutions}
%\footnote{AS: are these used to support any of our visions?}
%\subsection{Machine learning and genomic data}
We next present a conceptual model of the collaborative analysis of genomic data that summarises our visions.

\section{Conceptual system model}
\begin{figure}[htp]
 \centering
%\vspace*{-1.6cm}
\includegraphics[scale=0.62]{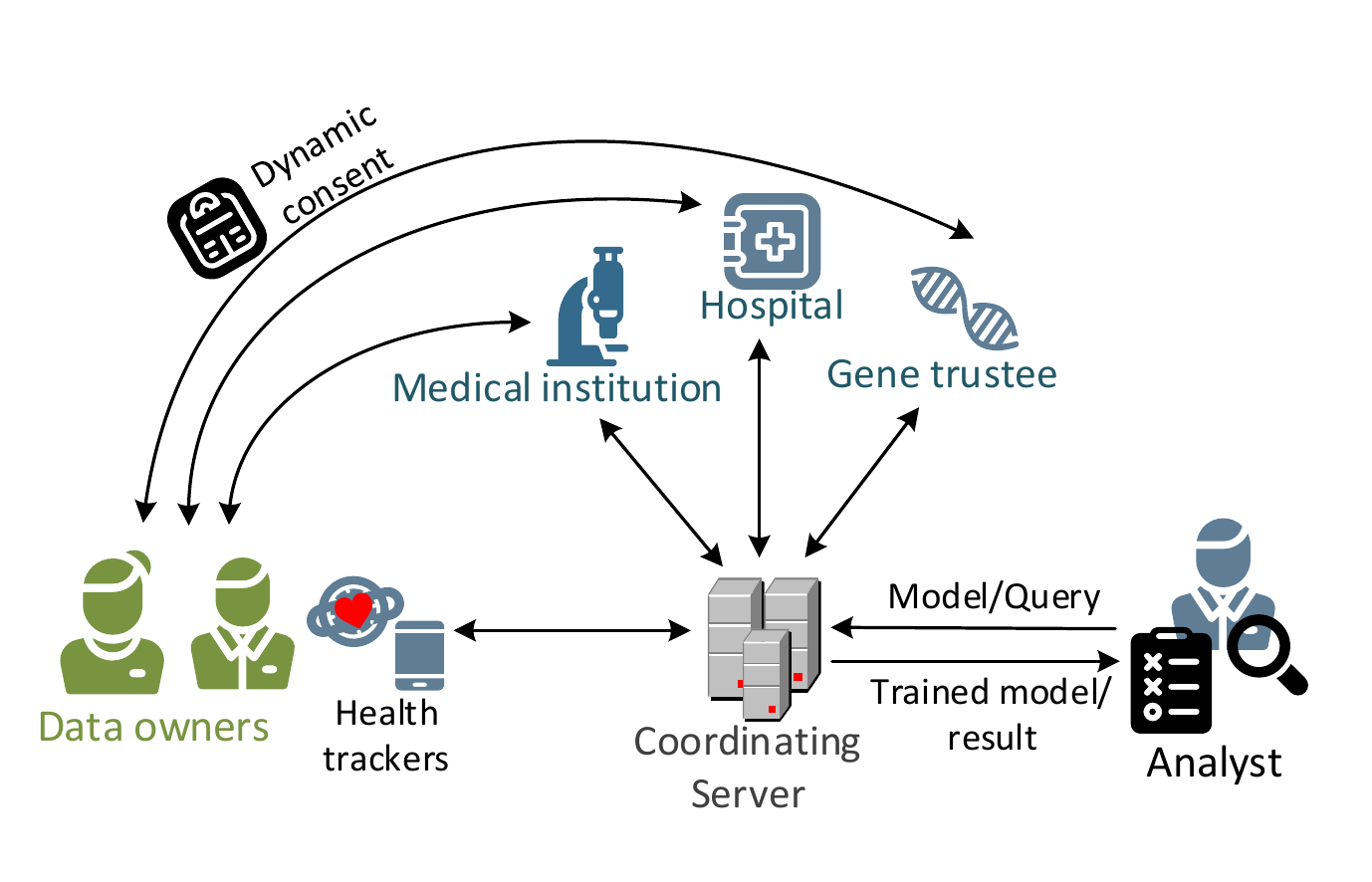}
\caption{Conceptual system model}
\label{fig:conceptual}
%\vspace{0.2cm}
\end{figure}

In a practical scenario and based on our visions, the data is stored in a distributed manner. Stored data can be genomic data, medical records, health data that comes from health trackers.
The conceptual system model in Fig. \ref{fig:conceptual} considers data is stored on all end devices/institutions, and data are encrypted while stored, processed, or in transit. In this conceptual high-level model, an analyst sends a request (that can be just a query or a model that needs to be trained). The coordinating server (by using the model-to-data method) sends the received model to the data holders, in which the consent has been given and is valid. Training a model can be done using federated learning and split learning techniques based on the computation resources and requirements. The query result can also be generated by using different encryption mechanisms that have special features. These techniques let the queries or computations be performed on encrypted data. The result of each of the requests is sent to the analyst in encrypted form.

It is worth mentioning that dynamic consent has been considered in this model; there is two-way communication between the data owner and data custodians to let individuals have control over the use of their data. %\footnote{AS: I do not understand this sentence.}
Moreover, considering the computation/query executions/model training are all done on the encrypted data, the secure/private processing and generate knowledge without sharing the data itself are satisfied. Different encryption techniques can be used to support this.
This way, by logging all the requests and results, transparency and accountability can be provided. 
Given the different nature of data stored in different data holders' storage, interconnection and integration can be provided if the requests/models need different data types as input and generate the outputs using this distributed information. It may require some repeats of a process and aggregation at the end to complete the request. Training a model or responding to a query using different distributed datasets, including genomic data and medical records needs further research.\par
We have designed a model, named PrivGenDB \cite{PrivGenDB}, that stores genomic data and other information of data owners in a secure way, and let the researchers/clinicians execute queries on the data in a secure and private way. This way, the data is not shared and just the result of queries is revealed to the authorised researchers/clinicians.
PrivGenDB does not address all the visions, e.g., it does not consider dynamic consent, and stores genomic data and other information of data owners centrally. However, the illustrated conceptual system model (Fig. \ref{fig:conceptual}) is more general, contains dynamic consent and it recommends having distributed storage. Therefore, a partial solution is proposed in \cite{PrivGenDB} but adapting the full system model presented here and have a solution for all of it is a possible future direction.

%We are interested in the visions elaborated in this article, and would like to work on our visions to establish them.

%--An example of this framework is presented in PrivGenDB

%This model has been partially adapted in PrivGenDB, but the model does not cover all the visions, or address all the open issues
%Solving them is a possible future direction

%PrivGenDB does not address dynamic consent. 

%A similar model was adapted 

%\footnote{AS: how PrivGenDB is related here? Is it an example of this model? Is it a simple use case of this? Or what? Finally, I do not get what is your point in the last two sentences.}
%show why you are interested in this vision and what you are doing about it.
\section{Open issues}
This section categorises and describes the open issues related to the discussed visions.
\subsubsection{Dynamic consent}
Consent may be approached in two ways, as previously mentioned: static consent and dynamic consent. Static consent necessitates approval for all future data uses at the time of data collection, which is usually done on paper. It can't answer problems that develop over time when the environment and requirements change, such as using the data for a different health initiative than the one for which consent was obtained. In this case, dynamic consent is beneficial. The data owner and data custodian have two-way communication, and the owner can update and give various forms of consent or revoke consent. %Furthermore, the data owner may oversee their health data usage over time and revoke consent. Moreover, the consent is transmitted with the related data when it is shared with third parties, and the participant has access to the research findings. Dynamic consent, on the other hand, has challenges, such as greater implementation costs, consent revocation, and data deletion assurance, as well as the requirement that patients have adequate digital expertise and time. 
Dynamic consent comes with its own set of problems, including higher implementation costs, consent revocation, and data deletion assurance, as well as the necessity that patients have sufficient technical knowledge and time.
Overall, the question of how to automate and manage consent in order to meet regulatory requirements, patient autonomy, cost, and data analytics is still an open issue.
\subsubsection{Genomic data with medical records}
Due to distributed nature of medical records and genomic data storage, it is essential to find a way to combine them for medical and also analysis uses. In the current situation, an individual's medical records are usually stored in hospital storage, and the genomic data is stored in a genome biobank or a gene trustee's storage. For training models, federated learning has already been proposed to develop the solution of training models in a distributed environment. However, there are no works on distributed different datasets that can be used for training a model. Even in the broader perspective, the health data that is collected from smart watches, mobile phones, etc., need also be interconnected with the medical records and genomic data. This is still an open issue that needs to be solved.
\subsubsection{Combining solutions}
Based on the existing challenges and barriers, different mechanisms have been studied and are going to be explored. One of the major concerns would be combining all these solutions on one platform to have all the visions altogether. This is challenging and needs the cooperation of experts from different fields, e.g., in providing genomic data privacy and security requirements, users (clinicians, analysts), biomedical engineers, computer engineers, etc. have to collaborate to be able to combine the solutions. Still, it is an open problem.

\subsubsection{Decentralisation}
In terms of decentralisation, for the better control of individuals over their own genomic data, it can be stored on their own devices. Therefore, anyone who wants to get access has to ask for permission. There are some suggestions of using blockchain and connecting the DNA data to the wallet of individuals, e.g., EncrypGen \cite{EncrypGen} that utilises blockchain and creates a platform so that users securely store and manage their DNA profiles and earn money by selling access to certain portions of the genome.
Decentralisation still has some problems and challenges, such as mobile/personal devices' storage and security, considering emergency situations, and if using blockchain, the blockchain security. Therefore, if considering full control over the data is an important factor, all these challenges must be considered and addressed.

%\section*{Acknowledgment}

\section{Conclusion}
This article presents our visions %\footnote{AS: you had more than a vision, right?}
of collaborative analysis of genomic data through the lens of moving from regulations and ethics towards utilising different techniques. Considering the sensitivity of genomic data, we discussed regulations and ethics that are currently being used. Then, through examining current detailed problems, our future visions are demonstrated, and some major existing techniques that can support them are listed. To provide an overview of our visions, a conceptual system model is demonstrated and explained. Finally, open issues are elaborated that need further study and research.

\remove{\section*{References}

Please number citations consecutively within brackets \cite{b1}. The 
sentence punctuation follows the bracket \cite{b2}. Refer simply to the reference 
number, as in \cite{b3}---do not use ``Ref. \cite{b3}'' or ``reference \cite{b3}'' except at 
the beginning of a sentence: ``Reference \cite{b3} was the first $\ldots$''

Number footnotes separately in superscripts. Place the actual footnote at 
the bottom of the column in which it was cited. Do not put footnotes in the 
abstract or reference list. Use letters for table footnotes.

Unless there are six authors or more give all authors' names; do not use 
``et al.''. Papers that have not been published, even if they have been 
submitted for publication, should be cited as ``unpublished'' \cite{b4}. Papers 
that have been accepted for publication should be cited as ``in press'' \cite{b5}. 
Capitalize only the first word in a paper title, except for proper nouns and 
element symbols.

For papers published in translation journals, please give the English 
citation first, followed by the original foreign-language citation \cite{b6}.}

\end{document}